\documentclass[12pt]{article}

\begin{document}

\title{\bf Layering and wetting transitions\break 
for an interface model}
\author{\large Salvador Miracle Sol\'e}
\date{\normalsize\it Centre de Physique Th{\'e}orique, 
CNRS, Marseille, France}
\maketitle

\pagestyle{myheadings}
\markright{\small\sl 
S.\ Miracle Sol{\'e} /  Layering and wetting transitions}

\begin{abstract}

We study the solid-on-solid interface model above a 
horizontal wall in three dimensional space, 
with an attractive interaction when the 
interface is in contact with the wall, at low temperatures. 
The system presents a sequence of layering transitions, 
whose levels increase with the temperature, 
before the complete wetting above a certain value of this quantity. 

\medskip

Published in                                                                                                                                                                                                                                                                                                                                                                                         
{\it Non-Equili\-brium Statistical Physics Today}, 
11th Gra\-nada Seminar, 
%13--17 September 2010,  
P.L. Garrido, J. Marro, F. de los Santos (Eds.), 
AIP Conference Proceedings, 
no.\ 1332, 
pp.\ 190--194
(ISBN 978-0-7354-0887-6),
American Institute of Physics, 
Melville, NY, 
2011. 

\end{abstract}

\noindent
{\footnotesize{\sc Keywords:}  SOS model, wetting and layering transitions, interfaces, entropic repulsion.}

\noindent
{\footnotesize{\sc Classification:} 68.08.Bc, 05.50.+q, 60.30.Hn, 02.50.-r}

%%%%%%%%%%%%%%%%%%%%%%%%%%%%%%%%%%%%%%%%%%%%
%% MAINMATTER
%%%%%%%%%%%%%%%%%%%%%%%%%%%%%%%%%%%%%%%%%%%%

\def\beq{\begin{equation}}
\def\eeq{\end{equation}}
\def\beqa{\begin{eqnarray}}
\def\eeqa{\end{eqnarray}}
\arraycolsep2pt

\bigskip

Consider the square lattice ${\bf Z}^2$.  
To each site $x=(x_1,x_2)$ of the lattice, 
an integer variable $\phi_x\ge 0$ is associated
which represents the height of the interface at this site. 
The system is first considered in a finite box $\Lambda\subset{\bf Z}^2$
with fixed values of the heights outside.
Each interface configuration on $\Lambda$: $\{\phi_x\}$, $x\in \Lambda$,
denoted $\phi_\Lambda$, has
an energy defined by the Hamiltonian
\beq 
H_\Lambda(\phi_\Lambda\mid\bar\phi)=2J
\sum_{\langle x,x'\rangle\cap \Lambda\ne\emptyset}
|\phi_x-\phi_{x'}|-2(J-K)\sum_{x\in \Lambda}\delta(\phi_x)+2J|\Lambda| ,
\label{H}\eeq
where $J$ and $K$ are positive constants,
the function $\delta$ equals $1$, when $\phi_x=0$, 
and $0$, otherwise, 
and $|\Lambda|$ is the number of sites in $\Lambda$. 
The first sum is taken over all nearest neighbors pairs 
$\langle x,x'\rangle\subset{\bf Z}^2$, such that at least one
of the sites belongs to $\Lambda$, and one takes $\phi_{x}=\bar\phi_{x}$
when $x\not\in \Lambda$,
the configuration $\bar\phi$ being the boundary condition,
assumed to be uniformly bounded. 

In the space ${\bf R}^3$, the region 
obtained as the union of all unit cubes centered at the sites 
of the lattice $\Lambda\times{\bf Z}$,  
that satisfy $x_3\le\phi(x_1,x_2)$, 
is supposed to be occupied by fluid $+$,  
while the complementary region above it, 
is occupied by fluid $-$. 
The common boundary between these regions
is a surface in ${\bf R}^3$, 
the microscopic interface.   
The region $x_3\le-1/2$ is considered as the substrate, also called the
wall $W$. 

The considered system differs from the usual SOS model
by the restriction 
to non-negative height
variables and the introduction of the second sum in
the Hamiltonian, the  
term describing the interaction with the substrate.

The probability of the configuration $\phi_\Lambda$,
at the inverse temperature $\beta=1/kT$,
is given by the finite volume Gibbs measure
\beq
\mu_\Lambda(\phi_\Lambda\mid\bar\phi)=Z(\Lambda,{\bar\phi})^{-1} 
\exp\big(-\beta H_\Lambda(\phi_\Lambda\mid \bar\phi)\big),
\label{mu}\eeq
where $Z(\Lambda,{\bar\phi})$ is the partition function
\beq
Z(\Lambda,\bar\phi)=\sum_{\phi_\Lambda} 
\exp\big(-\beta H_\Lambda(\phi_\Lambda\mid\bar\phi)\big).
\label{Xi}\eeq
Local properties at equilibrium can be described
by correlation functions between the heights on
finite sets of sites, obtained as expectations with respect 
to the Gibbs measure.

We next briefly describe some general results, 
which are an adaptation to our case
of analogous results established by Fr\"ohlich and Pfister
(ref.\  \cite{FP})
for the semi-infinite Ising model. 

Let $\Lambda\subset{\bf Z}^2$ be a rectangular box 
of sides parallel to the axes. 
Consider the boundary condition
$\bar\phi_x=0$, for all $x\not\in\Lambda$,
and write $Z(\Lambda,0)$ for the corresponding 
partition function.
The associated free energy per site,   
\beq 
\tau^{\scriptscriptstyle W-}=-\lim_{\Lambda\to\infty}(1/\beta|\Lambda|)
\ln Z(\Lambda,0),  
\label{tau}\eeq
represents the surface tension between the medium $-$ 
and the substrate $W$. 

This limit (\ref{tau}) exists and $0\le\tau^{\scriptscriptstyle W-}\le 2J$. 
One can introduce the densities  
\beq
\rho_z=\lim_{\Lambda\to\infty}\sum_{z'=0}^z
\langle \delta(\phi_x-z')\rangle^{(0)}_\Lambda,
\quad 
\rho_0=\lim_{\Lambda\to\infty}
\langle \delta(\phi_x)\rangle^{(0)}_\Lambda,
\eeq
Their connection with the surface free energy is given by the formula 
\beq
\tau^{\scriptscriptstyle W-}(\beta,K)=\tau^{\scriptscriptstyle W-}
(\beta,0)+2\int_0^K \rho_0(\beta,K')dK' .
\label{efe}\eeq

The surface tension $\tau^{\scriptscriptstyle W+}$
between the fluid $+$ and the substrate is $\tau^{\scriptscriptstyle W+}=0$. 
In order to define the surface tension $\tau^{\scriptscriptstyle +-}$
associated 
to a horizontal interface between the fluids $+$ and $-$ 
we consider the ordinary SOS model, 
with boundary condition ${\bar\phi}_x=0$.
The corresponding free energy gives $\tau^{\scriptscriptstyle +-}$.
With the above definitions, we have
\beq
\tau^{\scriptscriptstyle W+}(\beta)+ 
\tau^{\scriptscriptstyle +-}(\beta)\ge  
\tau^{\scriptscriptstyle W-}(\beta,K) .
\label{t} 
\eeq 
and the right hand side in (\ref{t}) is a monotone increasing and 
concave (and hence continuous) function of the parameter $K$.
This follows from relation (\ref{efe}) where the integrand
is a positive decreasing function of $K$.
Moreover, when $K\ge J$ equality is satisfied in (\ref{t}).

In the thermodynamic description of wetting, the partial wetting situation 
is characterized by the strict inequality in equation (\ref{t}),
which can occur only if $K<J$, as assumed henceforth. 
We must have then $\rho_0>0.$
The complete wetting situation is characterized by the equality in
(\ref{t}).
If this occurs for some $K$, say $K'<J$, then equation (\ref{efe})
tells us that this condition is equivalent to $\rho_0=0$.
Then, both conditions, the equality and $\rho_0=0,$ hold 
for any value of $K$ in the interval $(K',J)$. 

On the other hand, we know that $\rho_0=0$ implies also that 
$\rho_z=0$, for any positive integer $z$.
This indicates that, in the limit $\Lambda\to\infty$,
we are in the $+$ phase of the system, although 
we have used the zero boundary condition, so that the medium $-$
cannot reach anymore the wall. 
This means also that the Gibbs state of the SOS model does not exist
in this case.  

That such a situation of complete wetting is present for some 
values of the parameters does not follow, however, from the 
above results.
Actually this fact, as far as we know, remains an open problem
for the semi-infinite Ising model in 3 dimensions.
For the model we are considering an answer to this problem has
been given by Chalker \cite{Ch}.

\medskip

\noindent{\bf Chalker's result. }
{\it We use the following notation: 
\beq 
u=2\beta(J-K),\quad t=e^{-4\beta J}.
\label{ut}\eeq
If $u<-\ln(1-t^2)$, then $\rho_0=0$. } 

\medskip

Thus, for any given values of $J$ and $K$, 
there is a temperature below which the interface is almost surely bound and 
another higher one, above which it is almost surely unbound and  
complete wetting occurs. 
At low temperature (i.e., if $u>\ln16$), we have $\rho_0>0$.

The object of our study is to investigate the region not 
covered by these results when the temperature is low enough.
As mentioned in the abstract, we shall prove that 
a sequence of layering transitions
occurs before the system attains complete wetting.
More precisely the main results can be summarized as follows.

\medskip

\noindent{\bf Theorem 1. } 
{\it Let the integer $n\ge0$ be given. 
For any $\epsilon>0$ there exists a value $t_0(n,\epsilon)>0$ 
such that, if the parameters $t,u,$ satisfy 
$0<t<t_0(n,\epsilon)$ and
\beqa 
-\ln(1-t^2)+(2+\epsilon)t^{n+3}<&u&<-\ln(1-t^2)+(2-\epsilon)t^{n+2} , 
\label{Ass1}\eeqa
then the following statements hold:
(1) 
The free energy $\tau^{\scriptscriptstyle W-}$ is an
analytic function of the parameters $t,u$.
(2) 
There is a unique Gibbs state $\mu_n$, a pure phase associated to the
level $n$.
(3) 
The density is $\rho_0>0$. 
The second inequality in (\ref{Ass1}) is not needed in the case $n=0$. }

\medskip

An illustration for this theorem, in the plane $(K,\beta^{-1})$, 
is given in Figure 1.
From it we can see, as mentioned in the abstract, that if
the parameter $K$ is kept fixed, that seems natural 
since it depends on the properties of the substrate,
then the value $n$ of the level increases when the temperature 
is increased.

\medskip

Concerning this theorem, the following remarks can be made:

(1) The analyticity of the free energy comes from the existence
of a convergent cluster expansion for this system.
This implies the analyticity, in a direct way, of some
correlation functions and, in particular, of the density $\rho_0$.

(2) The unicity of the Gibbs state means that the 
correlation functions converge, when
$\Lambda\to\infty$, to a limit that does not depend on the chosen 
(uniformly bounded) boundary condition $\bar\phi_x$.
Being unique and translation invariant this state represents
a pure phase.
It is associated to a level $n$ 
in the sense that, for the typical configurations of the state,
large portions of the interface are near to the level $n$.

(3) The condition $\rho_0>0$ means that the interface remains at a 
finite distance from the wall and hence, we have partial wetting.
We can see that the region where this condition holds is, 
according to the Theorem, much larger, at low temperatures,
than the region initially proved by Chalker.
It comes very close to the line above which it is known 
that complete wetting occurs. 

(4) We have, 
$t_0(n,\epsilon)\to 0$ when $n\to\infty$ or $\epsilon\to0$.

\medskip

The reason why $t_0(n,\epsilon)$ depends on $\epsilon$,
satisfying remark 4, has an explanation.
One may believe that the regions of uniqueness of the state
extend in such a way that two %\break 
neighboring regions,
say those corresponding to the levels $n$ and $n+1$,
will have a common boundary where the two states $\mu_n$
and $\mu_{n+1}$ coexist.

%%%%%%%%%%%%%%%%
%%%                    Fig  1
%%%%%%%%%%%%%%%%%%%%%%%%%%%%

\setlength{\unitlength}{.7mm}

\bigskip\bigskip

\begin{center}
\begin{picture}(0,80)

\bezier{400}(30,70)(40,35)(40,0)

\bezier{400}(10,60)(40,30)(40,0)
\put(-13,60){\line(1,0){23}}
\put(3,53){\footnotesize $\mu_0$}

\bezier{400}(20,56)(40,28)(40,0)
\put(14,56){\line(1,0){6}}
\put(15,58){\footnotesize $\mu_1$}

\bezier{400}(28,53)(40,26)(40,0)
\put(22,53){\line(1,0){6}}
\put(22,55){\footnotesize $\mu_2$}

\bezier{400}(32,51)(40,25)(40,0)
\put(29,51){\line(1,0){3}}
\put(28.5,53){\footnotesize $\mu_3$}

\put(-30,-10){\vector(0,1){77}}
\put(-20,0){\vector(1,0){80}}

\put(-27,72){$\beta^{-1}$}
\put(64,-2){$K$}

\put(40,-2){\line(0,1){2}}
\put(38,-8){$J$}

\put(44,25){{\small wetting}}
\put(41,30){{\small complete}}
\end{picture}

\vskip1.5cm

{\small
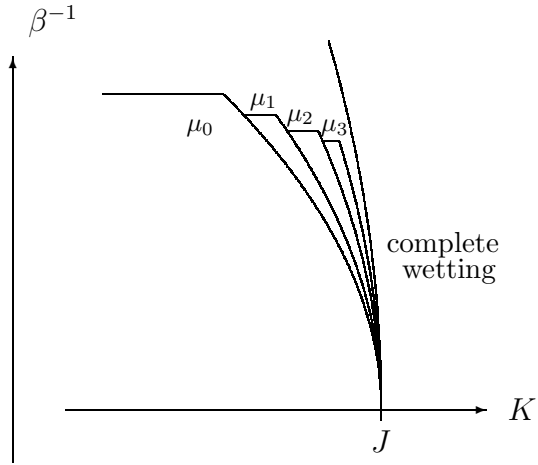
{\bf FIGURE 1. } The analyticity regions of Theorem 1. }

\end{center}

%%%%%%%%%%%%%%%%%%%%%%%%%%

\medskip

At this boundary
there will be a first order phase transition, since
the two Gibbs states are different.
The curve of coexistence does not exactly coincide
with the curve 
$u=-\ln(1-t^2)+2t^{n+3}$.
Theorem 1 says that it is however very near to it,
if the temperature is sufficiently low.

Let us formulate in the following statement the kind of theorem
that we expect. 
We think that such a statement could be proved using, as
for Theorem 1, an extension of the Pirogov-Sinai theory. 

\medskip

\noindent{\bf Statement. } 
{\it For each given integer $n\ge0$, there exists $t_0(n)>0$
and a continuous function $u=\psi_{n+1}(t)$ on the 
interval $0<t<t_0(n)$, such that 
the statements of Theorem 1 hold, 
for $t$ in this interval,   
in the region where $\psi_{n+1}(t)<u<\psi_{n+2}(t)$ and for $n=0$, 
in the region $\psi_{1}(t)<u$. 
When $u=\psi_{n+1}(t)$ the two Gibbs states, $\mu_n$ and $\mu_{n+1}$, 
coexist. }

\medskip

The existence of a sequence of
layering transitions has been proved for a related
model, known as the SOS model with an external magnetic field.
See the works by Dinaburg, Mazel \cite{DM}, 
Cesi, Martinelli \cite{CM} and Lebowitz, Mazel \cite{LM}.
This model has the same set of configurations as the model considered
here, but a different energy:
The second term in (\ref{H}) has to be
replaced by the term $+h\sum_{x\in\Lambda}\phi_x$ to obtain the
Hamiltonian of the model with an external magnetic field.
The method followed for the proof of the
Theorem is essentially analogous to the method
developed for the study of that model. 
The most important difference between the two systems
concerns the restricted ensembles and the computation
of the associated free energies.

Concerning the proof of Theorem 1 (paper in preparation) 
let us say that, 
for an interesting class of systems, among which our model is included, 
one needs some extension of the Pirogov-Sinai theory of phase transitions 
(see ref.\ \cite{Sin}). 
In such an  extension certain states, called the restricted ensembles, 
play the role of the ground states in the usual theory. 
They can be defined as a Gibbs probability measure
on certain subsets of configurations. 
In the present case one considers, for each $n=0,1,2,\dots$,   
subsets of configurations which are in some
sense  near to the constant configurations $\phi_x=n$. 

Namely, we consider the set ${\cal C}_k^{res}(\Lambda,n)$ 
of the microscopic interfaces, with boundary at height ${\bar\phi}_x=n$,
and whose Dobrushin walls have, all of them, horizontal projections with 
diameter less than $3k+3$ 
(these walls are the maximally connected sets of vertical plaquettes 
of the interface).
The Gibbs measure defined on the subset ${\cal C}_k^{res}(\Lambda,n)$ 
is the restricted ensemble corresponding to the level $n$.
The associated free energies per unit area  
\beq
f_k(n) = -\lim_{\Lambda\to\infty}({1/\beta{|\Lambda|}})\ln 
\sum_{\phi_\Lambda\in{\cal C}_k^{res}(\Lambda,n)}
\exp(-\beta H(\phi_\Lambda|n))
\eeq 
can be computed, with the help of cluster expansions 
(see, for instance, ref.\  \cite{M}), 
as a convergent power series in the variable $t$. 
Then one is able to study the phase diagram of the 
restricted ensembles. The restricted ensemble at level $n$
is said to be dominant, or stable, 
for some given values of the parameters $u$ and $t$, 
if $f_k(n)=\min_{n'}f_k(n')$. 
We then have: 

\medskip

\noindent{\bf Proposition 1. } 
{\it Let the integer $n\ge0$ be given and choose $k\ge 1$.
Let $a,b\ge0$ be two real numbers. Let $0<t\le t_1(k)=(3k+3)^{-4}$. If
\beq
-\ln(1-t^2)+(2+a)t^{n+3}\le u\le-\ln(1-t^2)+(2-b)t^{n+2} , 
\label{Ass}\eeq
then, we have
\beqa
& f_k(n) \le f_k(h)-at^{3n+3}+O(t^{3n+4}),\quad & \hbox{for any } h\ge n+1, 
\label{Pr1} \\
& f_k(n) \le f_k(h)-bt^{3n}+O(t^{3n+1}) ,\quad & \hbox{for any } 0\le h\le n-1. 
\label{Pr2} \eeqa }

\medskip

We notice that
$k\ge n$ is the useful case in the proof of Theorem 1, and
that the remainders in inequalities 
(\ref{Pr1}) and (\ref{Pr2}) can be bounded uniformly in $h$.
Then the proof of Theorem 1 consists in showing that the 
phase diagram of the pure phases at low temperature is close
to the phase diagram of the dominant restricted ensembles.

\end{document}